\documentclass[11pt]{article}

\usepackage[sort&compress,numbers]{natbib}
\usepackage{comment}


\usepackage{amssymb,amsmath}
\usepackage{graphicx}

\usepackage{url}
\urldef{\mailsa}\path|beka@cmu.edu, {sventura, msadinle, fienberg}@stat.cmu.edu|

\usepackage{arabtex}
\usepackage{utf8}
\usepackage{soul}

\usepackage[switch]{lineno}
\RequirePackage[colorlinks,citecolor=blue,urlcolor=blue]{hyperref}
\usepackage[ruled,lined]{algorithm2e}
\SetKw{KwSet}{Set}

\begin{document}

\title{A Comparison of Blocking \\Methods for Record Linkage}
\author{Rebecca C. Steorts and Cosma Rohilla Shalizi\\ Carnegie Mellon University}


\author{Rebecca C. Steorts, Samuel L. Ventura, \\ Mauricio Sadinle, Stephen E. Fienberg%
\thanks{This research was  partially supported by the National Science Foundation through grants SES1130706  and DMS1043903 to the Department of Statistics, Carnegie Mellon University.}%
 }

%

\maketitle

\vspace*{-1em}
\begin{abstract}
Record linkage seeks to merge databases and to remove duplicates when unique identifiers are not available.  Most approaches use blocking techniques to reduce the computational complexity associated with record linkage.  We review traditional blocking techniques, which typically partition the records according to a set of field attributes, and consider two variants of a method known as locality sensitive hashing, sometimes referred to as ``private blocking."  We compare these approaches in terms of their recall, reduction ratio, and computational complexity.   We evaluate these methods using different synthetic datafiles and conclude with a discussion of privacy-related issues.
\end{abstract}


\section{Introduction}
\label{sec:intro}
A commonly encountered problem in practice is merging databases
containing records  collected by different sources, often via
dissimilar methods.  Different variants of this task are known as record linkage, de-duplication, and
entity resolution.  Record linkage is inherently a difficult problem
\cite{christen_2011, Herzog_2007,Herzog:2010}.  
These difficulties are partially
due to the noise inherent in the data, which is often hard to
accurately model \cite{pasula_2003, steorts_2013b}.  A
more substantial obstacle, however,  is the scalability of the approaches \cite{WYP:2010}.  With $d$ databases of $n$ records each, brute-force approaches,
using all-to-all comparisons, require $O(n^d)$ comparisons.  This is quickly
prohibitive for even moderate $n$ or $d$.  To avoid this computational
bottleneck, the number of comparisons made must be drastically reduced, without
compromising linkage accuracy.  Record linkage is made scalable by ``blocking,'' which involves partitioning datafiles into
``blocks'' of records and treating records in different blocks as non-co-referent {\em a
  priori} \cite{christen_2011, Herzog_2007}. Record linkage methods are only
applied {\em within} blocks, reducing the comparisons to
$O(B n_{\max}^d)$, with $n_{\max}$ being the size of the largest of the $B$
blocks.

The most basic method for constructing a blocking partition picks certain fields (e.g. geography, or gender and year
of birth) and places records in the same block if and only if they agree on
all such fields.  This amounts to an {\em a priori} judgment that these fields
are error-free. We call this \emph{traditional blocking} (\S \ref{sec:block-naive}).

Other data-dependent blocking methods \cite{christen_2011, WYP:2010}
 are highly application-specific or are based on placing similar records into the
same block, using techniques of ``locality-sensitive hashing'' (LSH).
 LSH uses all of
the information contained in each record and can be adjusted to ensure that blocks are
manageably small, but then does not allow for further record linkage within blocks.  For example, \cite{christen_2014} introduced novel data structures for sorting and fast approximate nearest neighbor look-up within blocks produced by LSH.  Their
approach gave balance between speed and recall, but their technique is
very specific to nearest neighbor search with similarity defined by the hash
function.   Such
methods are fast and have high recall, but suffer from low precision, rather, too
many false positives.  This approach is called \emph{private} if, after the blocking is performed, all candidate records pairs are  compared and classified into matches/non-matches using computationally intensive ``private" comparison and classification techniques \cite{christen_2009}. 

Some blocking schemes involve clustering techniques to partition the records into clusters of similar records. \cite{mccallum_2000} used canopies, a simple clustering approach to group similar records into overlapping subsets for record linkage.  Canopies involves  organizing the data into overlapping clusters/canopies using an inexpensive distance measure.  Then a more expensive distance measure is used to link records within each canopy, reducing the number of required comparisons of records.
 \citep{vatsalan_2013} used a sorted nearest neighborhood clustering approach, combining $k$-anonymous clustering and the use of publicly available reference values to privately link records across multiple files. 

Such clustering-based blocking schemes motivate our variants of LSH methods for blocking. 
The first, transitive locality sensitive hashing (TLSH), is based upon the community discovery literature such that \emph{a soft transitivity} (or relaxed transitivity) can be imposed across blocks. The second, $k$-means locality sensitive hashing (KLSH), is based upon the information retrieval literature and  clusters similar records into blocks using a vector-space representation and projections.
 (KLSH has been
used before in information retrieval but never with record linkage
\citep{pauleve_2010}.)

The organization of this paper is as follows. \S \ref{sec:block} reviews traditional blocking. We then review other blocking  methods in \S \ref{sec:block-modern} stemming from the computer science literature. \S \ref{sec:lsh} presents two different methods based upon locality sensitive hashing, TLSH and KLSH.  We discuss the computational complexity of each approach in \S \ref{sec:complex}. We evaluate these methods (\S \ref{sec:results}) on simulated data using recall, reduction ratio, and the empirical computational time as our evaluation criteria, comparing to the other methods discussed above. Finally we discuss privacy protection aspects of TLSH and KLSH, given the description of LSH as a ``private" blocking technique. 

\section{Blocking Methods}
\label{sec:block}
Blocking divides records into mutually
exclusive and jointly exhaustive ``blocks,'' allowing the linkage
to be performed within each block. 
Thus, only records within the same
block can be linked; linkage algorithms may still aggregate information across
blocks. 
Traditional blocking requires domain knowledge to
pick out highly reliable, if not error-free, fields for blocking.  This methodology has at least two drawbacks.  The first is that the resulting blocks may still be so large that linkage within them is
computationally impractical.  The second is that because blocks {\em only}
consider selected fields, much time may be wasted comparing records that
happen to agree on those fields but are otherwise radically different.

We first review some simple alternatives to traditional blocking on fields, and then introduce
other blocking approaches that stem from computer science.

\subsection{Simple Alternatives to Blocking}
\label{sec:block-naive}
Since fields can be unreliable for many applications, blocking may miss large proportions of matches. Nevertheless, we can make use of domain-specific knowledge on the types of errors expected for field attributes. To make decisions about matches/non-matches, we must understand the \emph{kinds of errors} that are unlikely for a certain field or a combination of them.  With this information, we can identify a pair as a non-match when it has strong disagreements in a combination of fields.  It is crucial that this calculation be scalable since it must be checked for all pairs of records.  Some sequence of these steps reduces the set of pairs to a size such that more computationally expensive comparisons can be made.   In \S \ref{ss:naive_results}, we apply these concepts.

\subsection{Cluster-Based Blocking}
\label{sec:block-modern}
Others have described blocking as a clustering problem, 
sometimes with a special emphasis on privacy, e.g., see 
\cite{durham_2012,
karakasidis_2012, kuzu_2011, vatsalan_2013}.  
The motivation is
natural: the records in a cluster should be similar,
making good candidate pairs for linkage.

One clustering approach proposed for blocking is nearest neighbor clustering.
Threshold nearest neighbor clustering (TNN) begins with a single record as the base
of the first cluster, and recursively adds the nearest neighbors of records in
the cluster until the distance\footnote{The distance metric used can vary depending on the
nature of the records.} to the nearest neighbor exceeds some threshold.
Then one of the remaining records is picked to be the base for the next
cluster, and so forth.  K-nearest neighbor clustering (KNN) uses a similar procedure, but ensures
that each cluster contains at least $k$ records\footnote{Privacy-preserving versions of these approaches use ``reference
values'' rather than the records themselves to cluster the records \cite{vatsalan_2013}.}, to help maintain
``$k$-anonymity'' \cite{karakasidis_2012}. 
A major drawback of nearest neighbor clustering is that it requires
computing a large number of distances between records,  $O(n^2)$.  
Blocking a new record means finding its nearest neighbors, an
 $O(n)$ operation.  

The cost of calculating distances between records in
large, high-dimensional datasets led \cite{mccallum_2000} to propose the
method of \emph{canopies}.  In this approach, a computationally cheap (if inaccurate)
distance metric is used to place records into potentially-overlapping sets (canopies).  
An initial record is picked randomly to be the base of the first
canopy; all records within a distance $t_1$ of the base are grouped under that
canopy.  Those within distance $t_2 \leq t_1$ of the base are removed from
later consideration.  A new record is picked to be the base of the next
canopy, and the procedure is repeated until the list of candidate records is empty.  More accurate but
expensive distance measures are computed only between records that fall under
at least one shared canopy.  That is, only record-pairs sharing a canopy are candidates to be linked.


Canopies is not strictly a blocking method. They overlap, 
making the collection of canopies only a covering of the set
of records, rather than a partition.  We can derive blocks from canopies,
either set-theoretically or by setting $t_1=t_2$.  The complexity of building
the canopies is $O(nC_n)$, with $C_n$ being the number of canopies, itself a
complicated and random function of the data, the thresholds, and the order in
which records are chosen as bases. Further, finding fast, rough distance measures for complicated high-dimensional records is non-trivial.
%

\subsection{LSH-Based Approaches}
\label{sec:lsh}
We explore two LSH-based blocking methods.  These are based, respectively, on graph
partitioning or community discovery, and on combining random projections with
classical clustering.  The main reason for exploring these two methods is that even with comparatively
efficient algorithms for partitioning the similarity graph, doing that is still
computationally impractical for hundreds of thousands of records.

%

\subsubsection{Shingling}
LSH-based blocking schemes ``shingle''
\cite{rajaraman_2012} records.  That is, each record is treated as a string and
is replaced by a ``bag'' (or ``multi-set'') of length-$k$ contiguous
sub-strings that it contains. These are known as ``$k$-grams'', ``shingles'',
or ``tokens''.  For example, the string ``TORONTO'' yields the bag of length-two
shingles ``TO'', ``OR'', ``RO'', ``ON'', ``NT'', ``TO''.  (N.B., ``TO'' appears
twice.)

As alternative to shingling, we might use a bag-of-words representation, or
even to shingle into consecutive pairs (triples, etc.) of words. 
In our
experiments, shingling at the level of letters worked better than dividing by
words.

\subsubsection{Transitive LSH (TLSH)}
\label{subsec:tlsh}

We create a graph of the similarity between records.  
For simplicity, assume that all fields are string-valued.  Each record is
shingled with a common $k$, and the bags of shingles for all $n$ records are
reduced to an $n$-column binary-valued matrix $M$, indicating which
shingles occur in which records.  
$M$ is large, since the number of length-$k$ shingles typically grows
exponentially with $k$.  As most shingles are absent from most records, $M$ is
 sparse.  We reduce its dimension by generating a random
``minhash'' function and applying it to each column.  Such functions map
columns of $M$ to integers, ensuring that the probability of two columns being
mapped to the same value equals the Jaccard similarity between the columns
\cite{rajaraman_2012}.  Generating $p$ different minhash functions, we reduce
the large, sparse matrix $M$ to a dense $p\times n$ matrix, $M^{\prime}$, of integer-valued
``signatures,'' while preserving information.  Each row of
$M^{\prime}$ is a random projection of $M$.  Finally, we divide
the rows of $M^{\prime}$ into $b$ non-overlapping ``bands,'' apply a hash
function to each band and column, and establish an edge between two records if
their columns of $M^{\prime}$ are mapped to the same value in any
band.\footnote{To be mapped to the same value in a particular band, two columns
  must either be equal, or a low-probability ``collision'' occurred for the
  hash function.}

These edges define a graph: records are nodes, and edges indicate a certain
degree of similarity between them. We form blocks by dividing the graph into its connected components.
However, the largest connected
components are typically very large, making them unsuitable as blocks.
Thus, we  sub-divide the connected components into ``communities'' or ``modules'' 
--- sub-graphs that are densely connected
internally, but sparsely connected to the rest of the graph. This 
ensures that the blocks  produced consist of records that are all highly
similar, while having relatively few ties of similarity to
records in other blocks \cite{fortunato_2010}.  Specifically, we apply the
algorithm of \cite{clauset_2004}\footnote{We could use other community-discovery algorithms, e.g. \cite{goldenberg_2010}.}, sub-dividing communities greedily, until even
the largest community is smaller than a specified threshold.\footnote{This maximum
size ensures that record linkage is feasible.}  
 The end result is a set of blocks that balance false
negative errors in linkage (minimized by having a few large blocks) and the
speed of linkage (minimized by keeping each block small).  We summarize the
whole procedure in Algorithm \ref{subsec:tlsh} (see Appendix \ref{sec:app}).

TLSH involves many tuning parameters (the length of shingles, the number
of random permutations, the maximum size of communities, etc.)
We chose the shingle such that we have 
the highest recall possible for each application. We used a random permutation of 100, since the recall was approximately constant for all permutations higher than 100. 
Furthermore, we chose a maximum size of the communities of 500, after tuning this specifically for desired speed.


\subsubsection{K-Means Locality Sensitive Hashing (KLSH)}
\label{subsec:klsh}

The second LSH-based blocking method begins, like TLSH,
by shingling the records, treated as strings, but then differs in several ways.
First, we do not ignore the number of times each shingle type appears in a
record, but rather keep track of these counts, leading to a bag-of-shingles
representation for records.  Second, we measure similarity between
records using the inner product of bag-of-shingles vectors, with
inverse-document-frequency (IDF) weighting. Third, we reduce the
dimensionality of the bag-of-shingles vectors by random projections, followed
by clustering the low-dimensional projected vectors with the $k$-means
algorithm. 
Hence, we can control the mean
number of records per cluster to be $n/c$, where $c$ is the number of block-clusters.  In practice, there is a fairly small
dispersion around this mean, leading to blocks that, by construction, have the roughly the same distribution for all applications.\footnote{This property is not guaranteed for most LSH methods.}  The KLSH algorithm is given in Appendix \ref{sec:app}.

\section{Computational Complexity}
\label{sec:complex}

\subsection{Computational Complexity of TLSH}

The first steps of the algorithm can be done independently across records.
Shingling a single record is  $O(1),$  so shingling all the records
is $O(n)$.  Similarly, applying one minhash function to the shingles of one
record is $O(1),$ and there are $p$ minhash functions, so minhashing
takes $O(np)$ time.  Hashing again, with $b$ bands, takes $O(nb)$ time.  We
 assume that $p$ and $b$ are both $O(1)$ as $n$ grows.


We create an edge between every pair of records that get mapped to the same
value by the hash function in some band.  Rather than iterating over pairs of
records, it is faster to iterate over values $v$ in the range of the
hash function.  If there are $|v|$ records mapped to the value $v$, creating
their edges takes $O(|v|^2)$ time. On average, $|v| = n V^{-1}$, where $V$ is the
number of points in the range of the hash function, so creating the edge list
takes $O(V (n/V)^2 ) = O(n^2 V^{-1})$ time.  \cite{clauset_2004} shows that creating
the communities from the graph is $O(n (\log{n})^2)$.


The total complexity of TLSH is $O(n) + O(np) + O(nb) + O(n^2 V^{-1}) +
O(n(\log{n})^2) = O(n^2 V^{-1})$, and is dominated by actually building the graph.


\subsection{Computational Complexity of KLSH}

As with TLSH, the shingling phase of KLSH takes $O(n)$ time.  The time required
for the random projections, however, is more complicated.  Let $w(n)$ be the
number of distinct words found across the $n$ records.  The time needed to do
one random projection of one record is then $O(w(n))$, and the time for the
whole random projection phase is $O(npw(n))$.  For $k$-means cluster, with a
constant number of iterations $I$, the time required to form $b$ clusters of
$n$ $p$-dimensional vectors is $O(bnpI)$.  Hence, the complexity is $O(npw(n))
+ O(bnpI)$.


Heaps's law suggests $w(n) = O(n^\beta)$, where $0 < \beta
< 1$.\footnote{For English text, $0.4 < \beta < 0.6$.}  Thus, the complexity
is $O(p n^{1+\beta}) + O(bnpI)$.  
For record linkage to run in linear time, it must
run in constant time in each block. Thus, the number of records per block must be
constant, i.e., $b = O(n)$. Hence, the time-complexity for blocking
is $O(p n^{1+\beta}) + O(n^2 pI) = O(n^2 pI),$ a quadratic time algorithm
dominated by the clustering.  Letting $b = O(1)$ yields an over-all time
complexity of $O(p n^{1+\beta})$, dominated by the projection step.  If we
assume $\beta = 0.5$ and let $b = O(\sqrt{n}),$  then both the projection and
the clustering steps are $O(pn^{1.5})$.  Record linkage in each block is $O(n),$ so record linkage is $O(n^{1.5}),$ 
rather than $O(n^2)$ without blocking.

\subsection{Computational Complexity of Traditional Blocking Approaches}
Traditional blocking approaches use attributes of the records to partition records into blocks.  As such, calculating the blocks using traditional approaches requires $O(n)$ computations.  For example, approaches that block on birth year only require a partition of the records based on these fields.  That is, each record is simply mapped to one of the unique birth year values in the dataset, which is an $O(n)$ calculation for a list of size $n$. Some traditional approaches, however, require $O(n^2)$ computations.  For example, in Table \ref{t:naive_results}, we show some effective blocking strategies which require $O(n^2)$ computations, but each operation is so cheap that they can be run in reasonable time for moderately sized files.

\section{Results}
\label{sec:results}
We test the previously mentioned approaches on data from the RecordLinkage R package.\footnote{\url{http://www.inside-r.org/packages/cran/RecordLinkage/docs/RLdata}}
These simulated datasets contain 500 and 10,000 records (denoted \texttt{RLdata500} and \texttt{RLdata10000}), with exactly 10\% duplicates in each list. These datasets contain first and last Germanic name and full date of birth (DOB). Each duplicate contains one error with respect to the original record, and there is maximum of one duplicate per original record.  Each record has a unique identifier, allowing us to test the performance of the blocking methods. 

We explore the performance of the previously presented methods under other scenarios of measurement error.  \citep{Christen05, ChristenPudjijono09, ChristenVatsalan13} developed a data generation and corruption tool that creates synthetic datasets containing various field attributes.  This tool includes  dependencies between fields and permits the generation of different types of errors. We now describe the characteristics of the datafiles used in the simulation.  We consider three files having the following field attributes: first and last name, gender, postal code, city, telephone number, credit card number, and age. For each database, we allow either 10, 30, or 50\% duplicates per file, and each duplicate has five errors with respect to the original record, where these five errors are allocated at random among the fields.  Each original record has maximum of five duplicates.  We refer to these files as the ``noisy'' files.


%
\subsection{Traditional Blocking Approaches}\label{ss:naive_results}
Tables \ref{t:naive_results} -- \ref{t:naive_results2} provide results of traditional blocking when applied to the \texttt{RLdata10000} and ``noisy" files.  While field-specific information \emph{can} yield favorable blocking solutions,  each blocking criteria is application specific. The overall goal of blocking is to reduce the overall set of candidate pairs, while minimizing the false negatives induced. Thus, we find the \emph{recall} and \emph{reduction ratio} (RR). This corresponds to the proportion of true matches that the blocking criteria preserves, and the proportion of record-pairs  discarded by the blocking, respectively.   

Criteria 1 -- 5 (Table \ref{t:naive_results}) and 1 -- 6 (Table \ref{t:naive_results2}) show that \emph{some} blocking approaches are poor, where the recall is never above 90\%.  Criteria requiring exact agreement in a single field or on a combination of them are  susceptible to field errors.  More reliable criteria are constructed using combinations of fields such that multiple disagreements must be met for a pair to be declared as a non-match.  (See Criteria 7--10 and 12 in Table \ref{t:naive_results}, and 7 -- 8 in Table \ref{t:naive_results2}.) We obtain high recall and  RR using these, but in general their performance is context-dependent.  

Criteria 10 (Table \ref{t:naive_results}) deals with the case when a pair is declared a non-match whenever it disagrees in four or more fields, which is reliable since false-negative pairs are only induced when the datafile contains large amounts of error.  For example this criterion does not lead to good results with the noisy files, hence a stronger criteria is needed, such as 7  (Table \ref{t:naive_results2}).  Using Criteria 12 (Table \ref{t:naive_results}) and 8 (Table \ref{t:naive_results2}), we further reduce the set of candidate pairs whenever a pair has a strong disagreement in an important field.\footnote{We use the Levenshtein distance (LD) of first and last names for pairs passing Criterion 10 of Table \ref{t:naive_results} or Criteria 7 of Table \ref{t:naive_results2}, and declare  pairs   as non-matches when LD $\geq 4$ in either first or last name.} These criteria are robust. In order to induce false negatives, the error in the file must be much higher than expected.

\begin{table}[htdp]
\begin{center}
\begin{tabular}{rlrr}
				\hline\\[-8pt]
				& Declare non-match if disagreement in: & Recall (\%) & RR (\%) \\
				\hline\\[-8pt]
				1.&First OR last name & 39.20 & 99.98\\
				2.&Day OR month OR year of birth & 59.30 & 99.99\\
				3.&Year of birth & 84.20 & 98.75\\
				4.&Day of birth & 86.10 & 96.74\\
				5.&Month of birth & 88.40 & 91.70\\
				6.&Decade of birth & 93.20 & 87.76\\
				7.&First AND last name & 99.20 & 97.36 \\
				8.&\{First AND last name\} OR &&\\
				&\{day AND month AND year of birth\} & 99.20 & 99.67 \\
				9.&Day AND month AND year of birth & 100.00 & 87.61\\
				10.&More than three fields & 100.00 & 99.26 \\
				11.&Initial of first OR last name & 100.00 & 99.25\\
				12.&\{More than three fields\} OR &&\\
				& \{Levenshtein dist. $\geq 4$ in first OR last name\}& 100.00 & 99.97\\
				\hline
							\end{tabular}
							
\end{center}
 \caption{Criteria for declaring pairs as non-matches, where results correspond to the \texttt{RLdata10000} datafile.  }
 \label{t:naive_results}

\end{table}%


\begin{table}[htdp]
\begin{center}
\begin{tabular}{rlrr}
  \hline\\[-8pt]
	& Declare non-match if disagree in: & Recall (\%) & RR (\%) \\
  \hline\\[-8pt]
	1.&Gender                      &	  31.96	&	 53.39\\
	2.&City                         &	  31.53	&	 77.25\\
	3.&Postal Code                 &	  32.65	&	 94.20\\
	4.&First OR last name          &	   1.30	&	$>$99.99\\
	5.&Initial of first OR last name&	  78.10	&	 99.52\\
	6.&First AND last name          &	  26.97	&	 99.02\\
	7.&All fields                 &	    93.28	&	 40.63   \\
	8.&\{All fields\} OR \{Levenshtein dist. &&\\
	&   $\geq 4$ in first OR last name\} &	 92.84	&	 99.92\\
	\hline
\end{tabular}
\end{center}
 \caption{Criteria for declaring pairs as non-matches, where results correspond to the noisy datafile with 10\% duplicates.  Similar results obtained for 30 and 50\% duplicates.  }
 \label{t:naive_results2}
\end{table}%

\vspace*{-2em}

\subsection{Clustering Approaches}
\label{sec:cluster-results}

\begin{figure}[ht]
\centering
\includegraphics[width=0.48\textwidth]{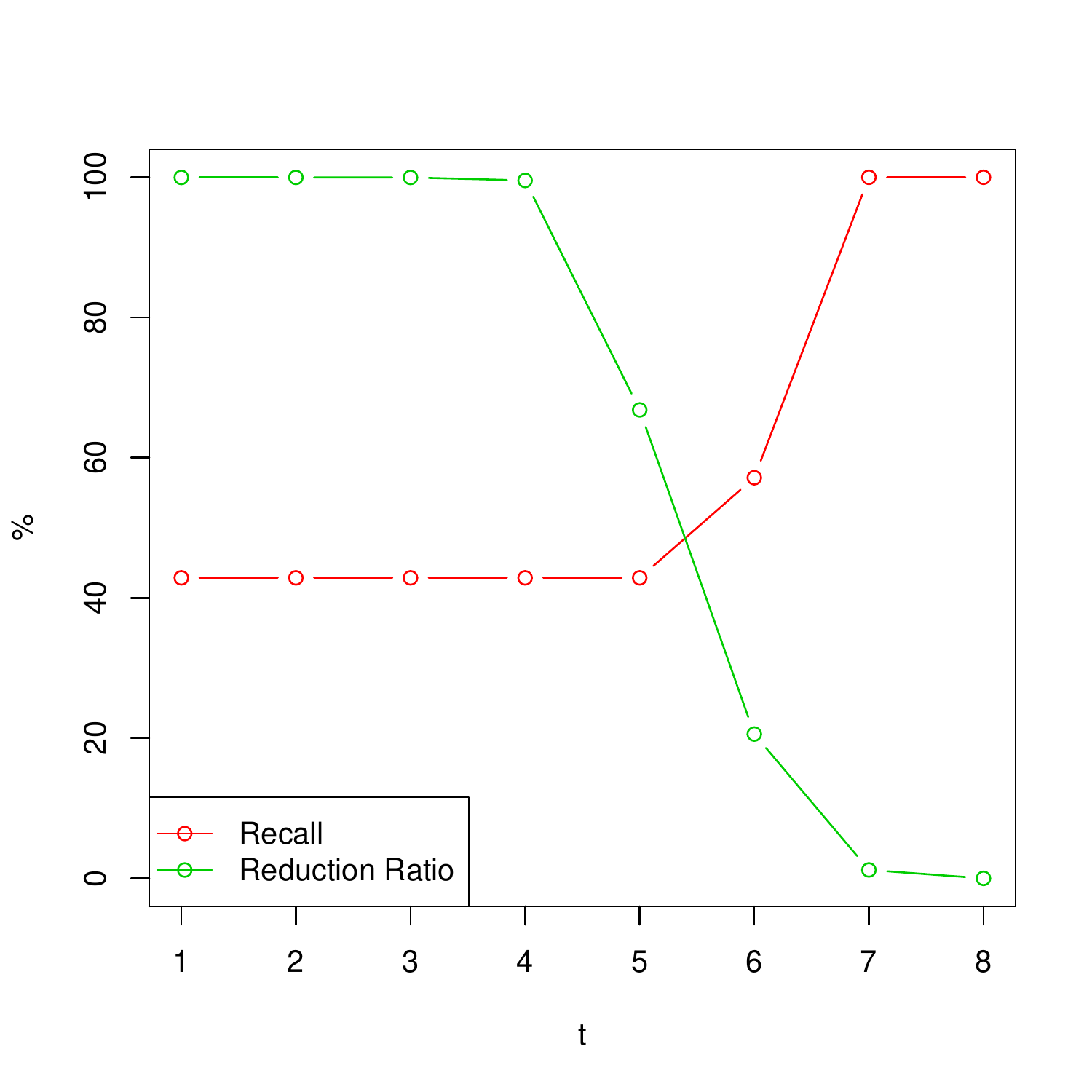} \includegraphics[width=0.48\textwidth]{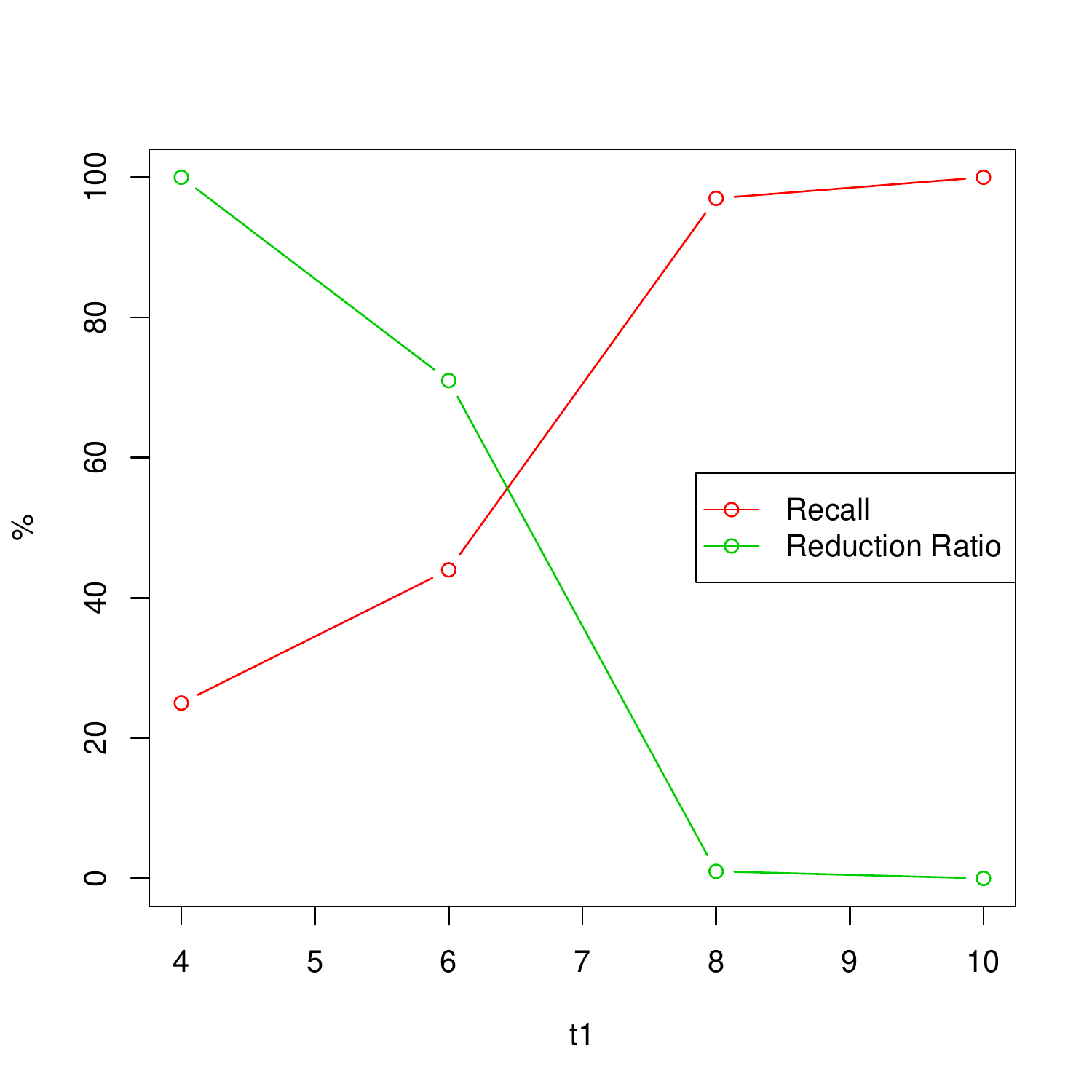}
\caption{Performance of threshold nearest neighbors (left) and canopies (right) on the RLdata10000 datafile.}
\label{TNN_and_canopies}
\end{figure}

Our implementations of \cite{mccallum_2000}'s canopies approach and \cite{vatsalan_2013}'s nearest neighbor approach perform poorly on the \texttt{RLdata10000} and ``noisy" datasets\footnote{In our implementations, we use the TF-IDF matrix representation of the records and Euclidean distance to compare pairs of records in TNN and canopies.  We tried several other distance measures, each of which gave similar results.}. Figure \ref{TNN_and_canopies} gives results of these approaches for different threshold parameters ($t$ is the threshold parameter for sorted TNN) for the \texttt{RLdata10000} dataset.  For all thresholds, both TNN and canopies fail to achieve a balance of high recall and a high reduction ratio.

%
%

\begin{figure}[ht]
\centering
\includegraphics[width=0.48\textwidth]{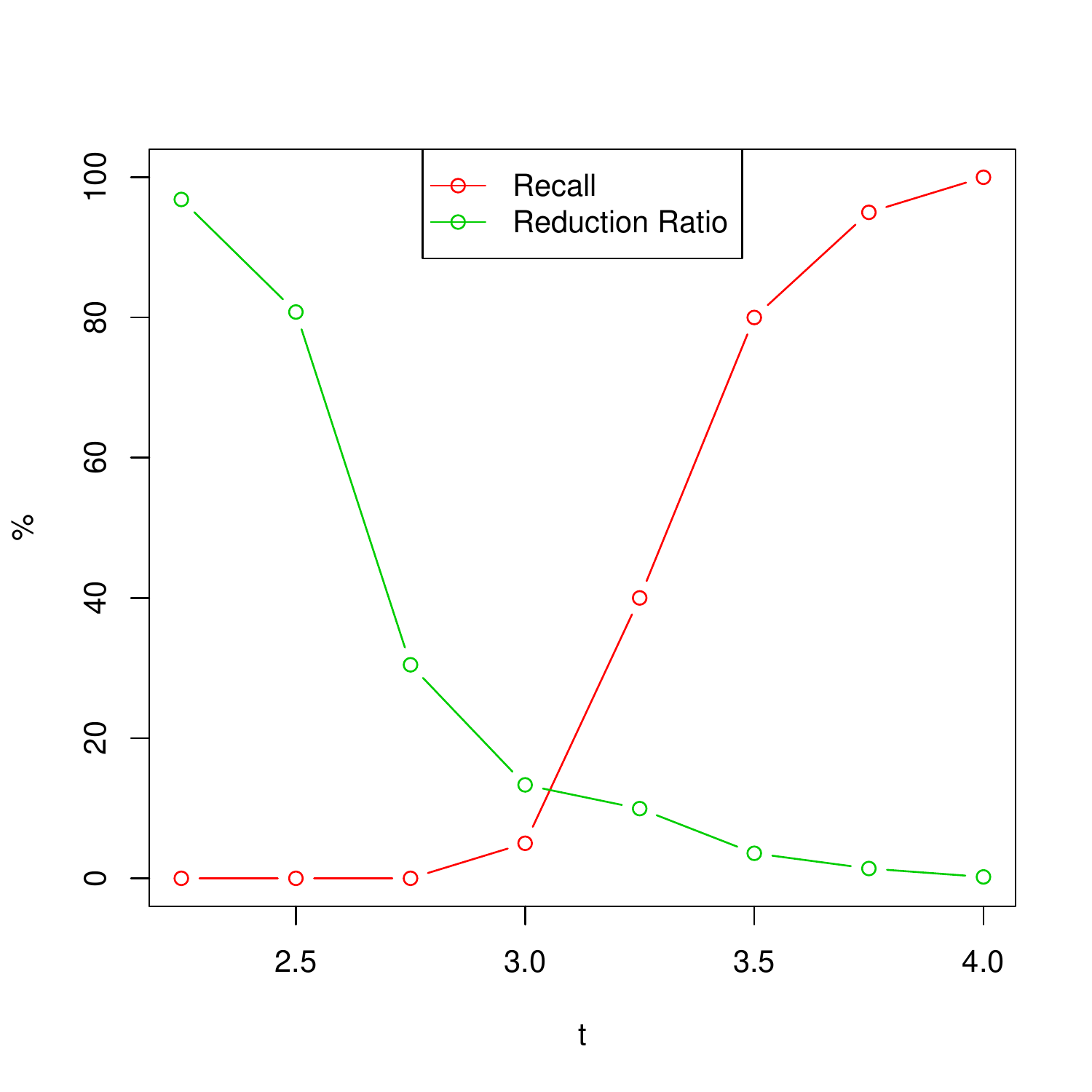} \includegraphics[width=0.48\textwidth]{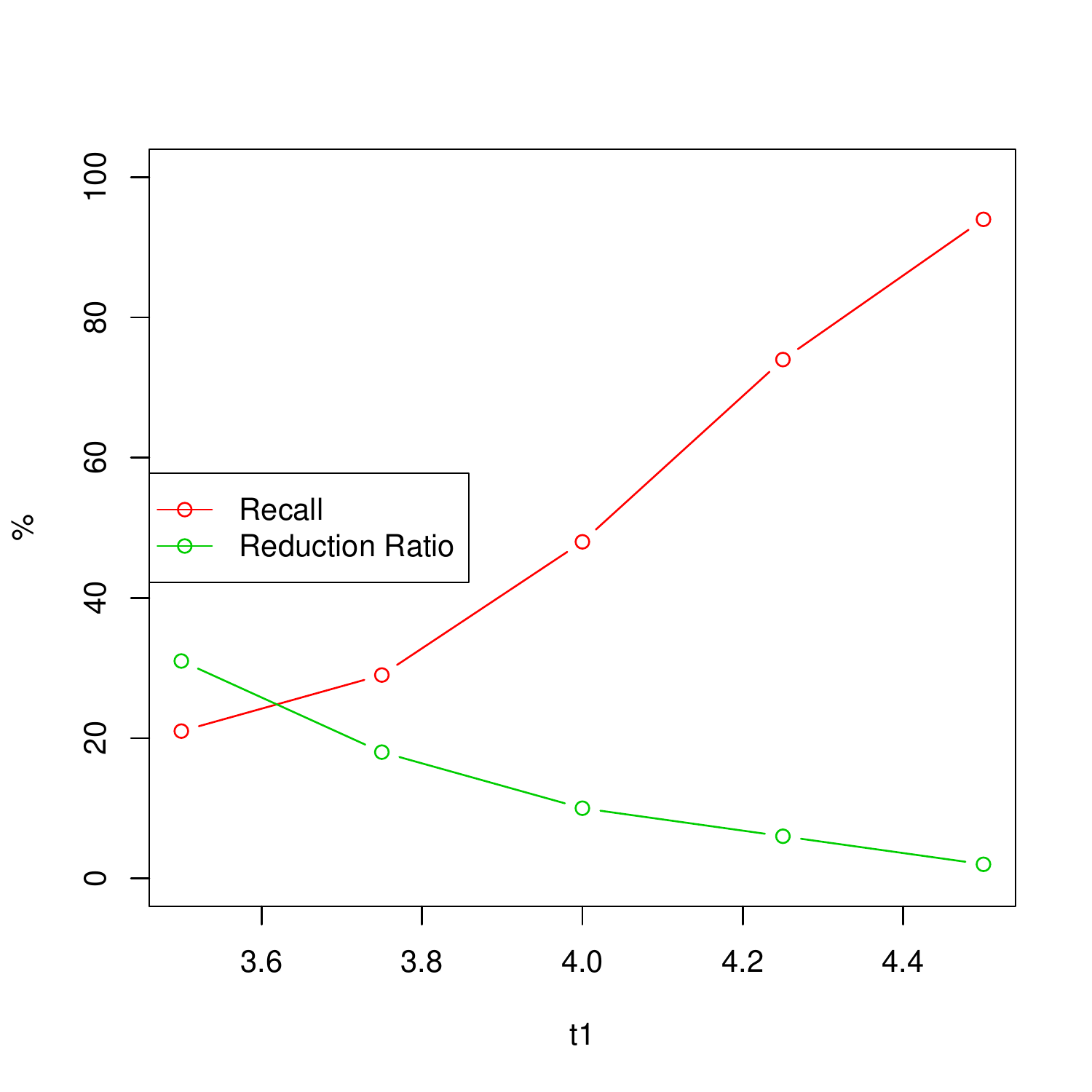}
\caption{Performance of TNN (left) and canopies (right) on the ``noisy" datafile (10\% duplicates).  The other ``noisy" datafiles exhibited similar behavior as the figures above.  }
\label{TNN_and_canopies_error}
\end{figure}

Turning to the ``noisy" dataset with 10\% duplicates, we find that TNN fails to achieve a balance of high recall and high reduction ratio, regardless of the threshold $t$ that is used.  Similarly, the canopies approach does not yield a balance of high recall while reducing the number of candidate pairs.

Clearly, both clustering approaches fail to achieve a balance of high recall and RR for any threshold parameters.  The inefficacy of these approaches is likely due the limited number of field attributes (five fields) and the Euclidean distance metric used for these datasets.  In particular, only three fields in the ``noisy" dataset use textual information, which both of these approaches use to identify similar records.  Limited field information can make it difficult for clustering approaches to group similar records together, since the resulting term frequency matrices will be very sparse.  Thus, we investigate the behavior with the same number of duplicates, but vary the error rate and provide richer information at the field attribute level. Figure \ref{TNN_and_canopies_error} illustrates that both methods do not have a good balance between recall and RR, which we investigated for various thresholds.  As such, further analysis of these approaches on more information-rich datasets is required in order to make sound conclusions about their efficacy for blocking. (We note that the metrics used in TLSH and KLSH, which shingle the records, were chosen so as to not have such problems.) 


\subsection{LSH Approaches}
\label{sec:data500}
Since the  performance of KLSH and TLSH depends on tuning parameters, we tune each application appropriately to these.  We empirically measure the scalability of these methods, which are consistent with our derivations in \S \ref{sec:complex}.

\begin{figure}
\centering
\includegraphics[width=0.45\textwidth]{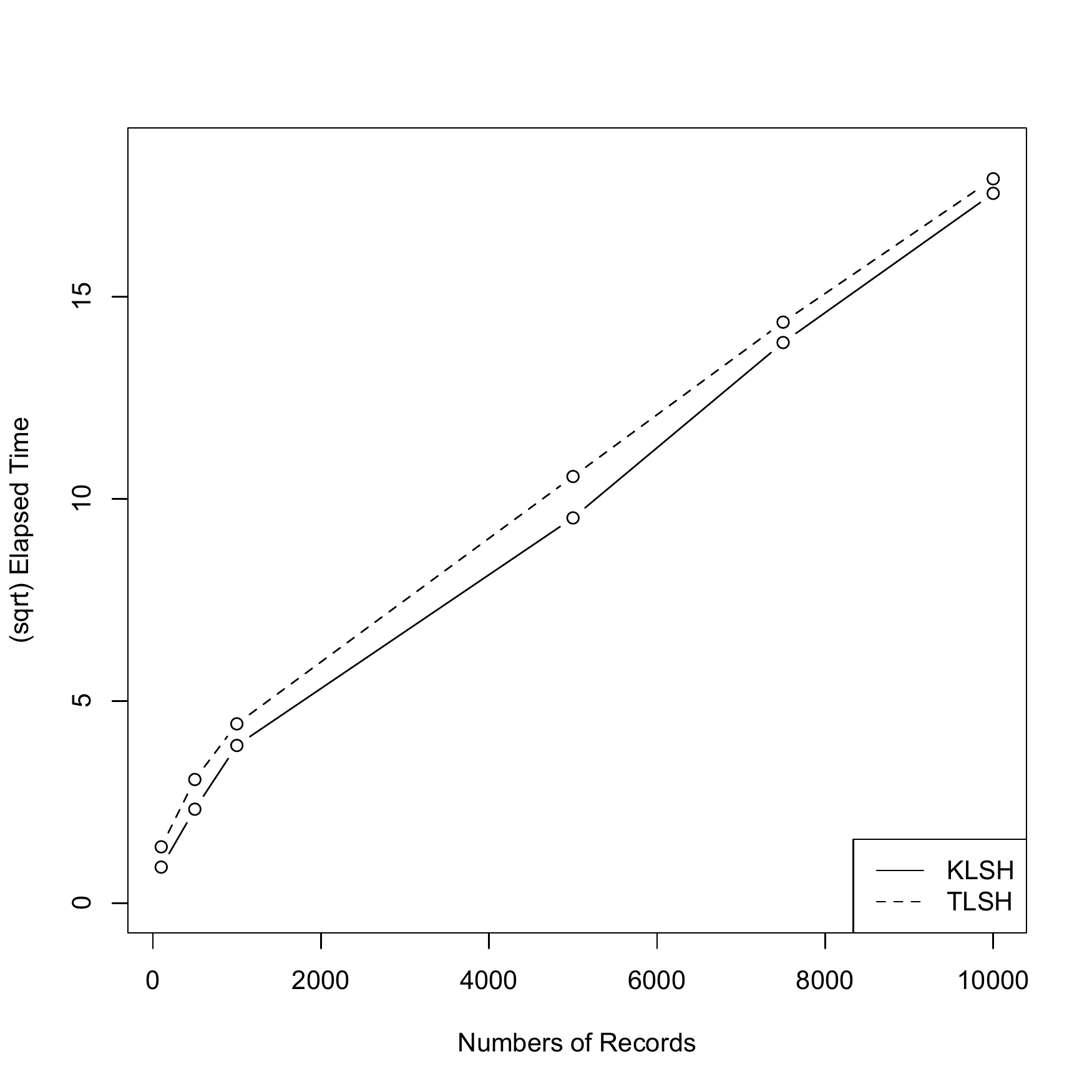}
\includegraphics[width=0.45\textwidth]{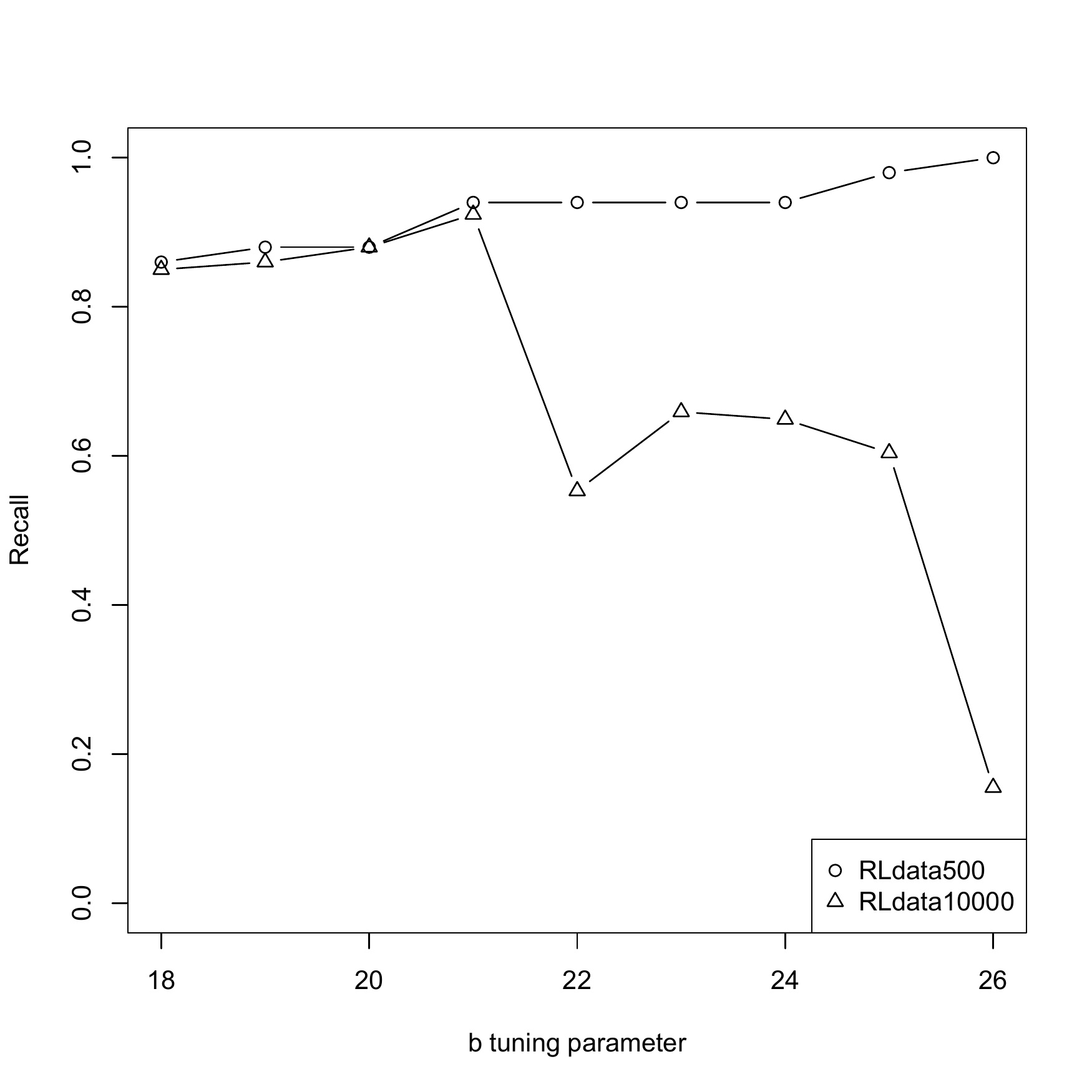}
\caption{\text{RLdata10000/RLdata500} datasets. Left: Square Root Elapsed time versus number of records for KLSH and TLSH, illustrating that both methods scale nearly quadratically (matching the computationally complexity findings).  We shingle using $k=5$ for both methods.  We use a band of 26 for TLSH.  Right: Recall versus $b$ for both \texttt{RLdata500} and \texttt{RLdata10000} after running TLSH.}
\label{comp_time}
\end{figure}

\begin{figure}
\centering
\includegraphics[width=0.45\textwidth]{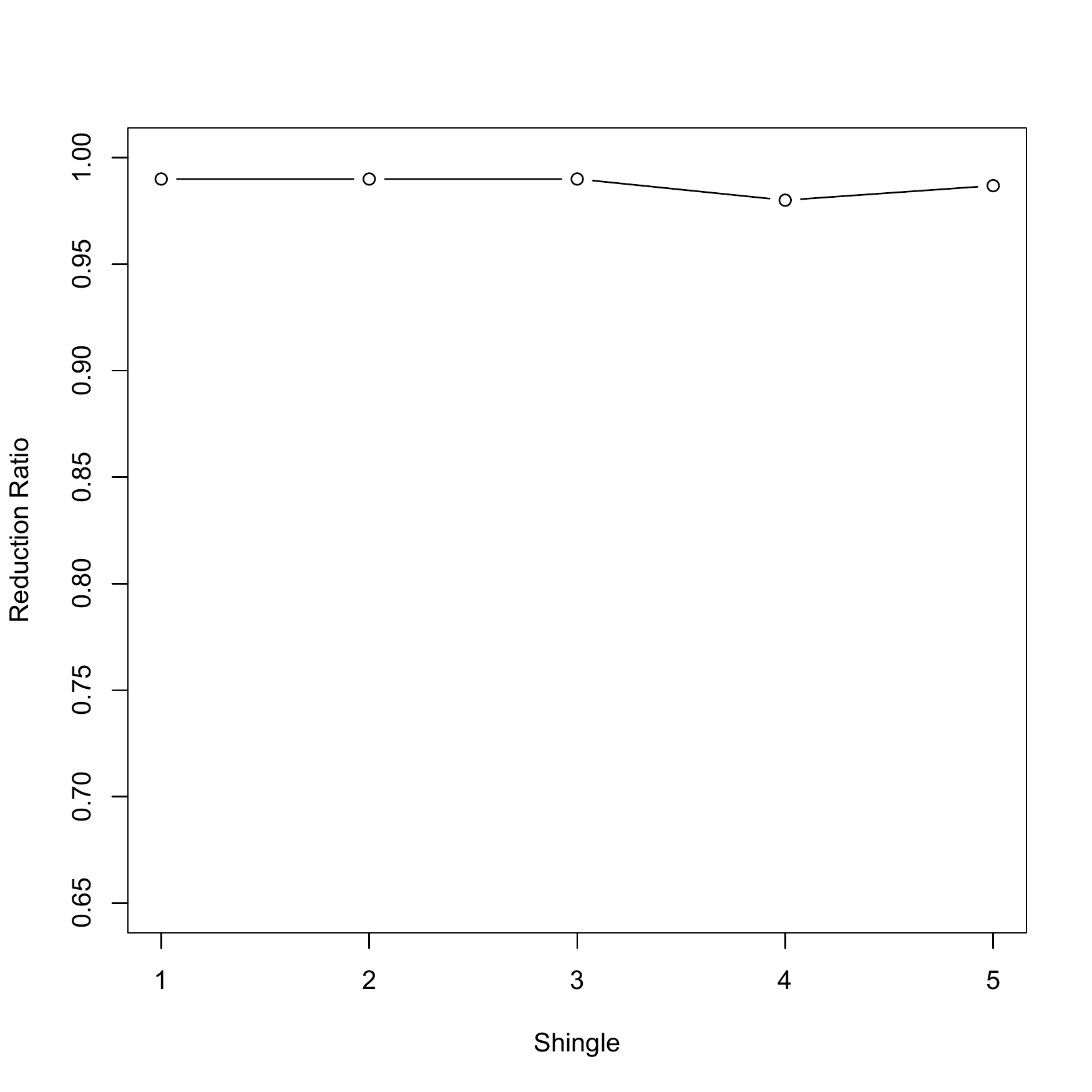}
\includegraphics[width=0.45\textwidth]{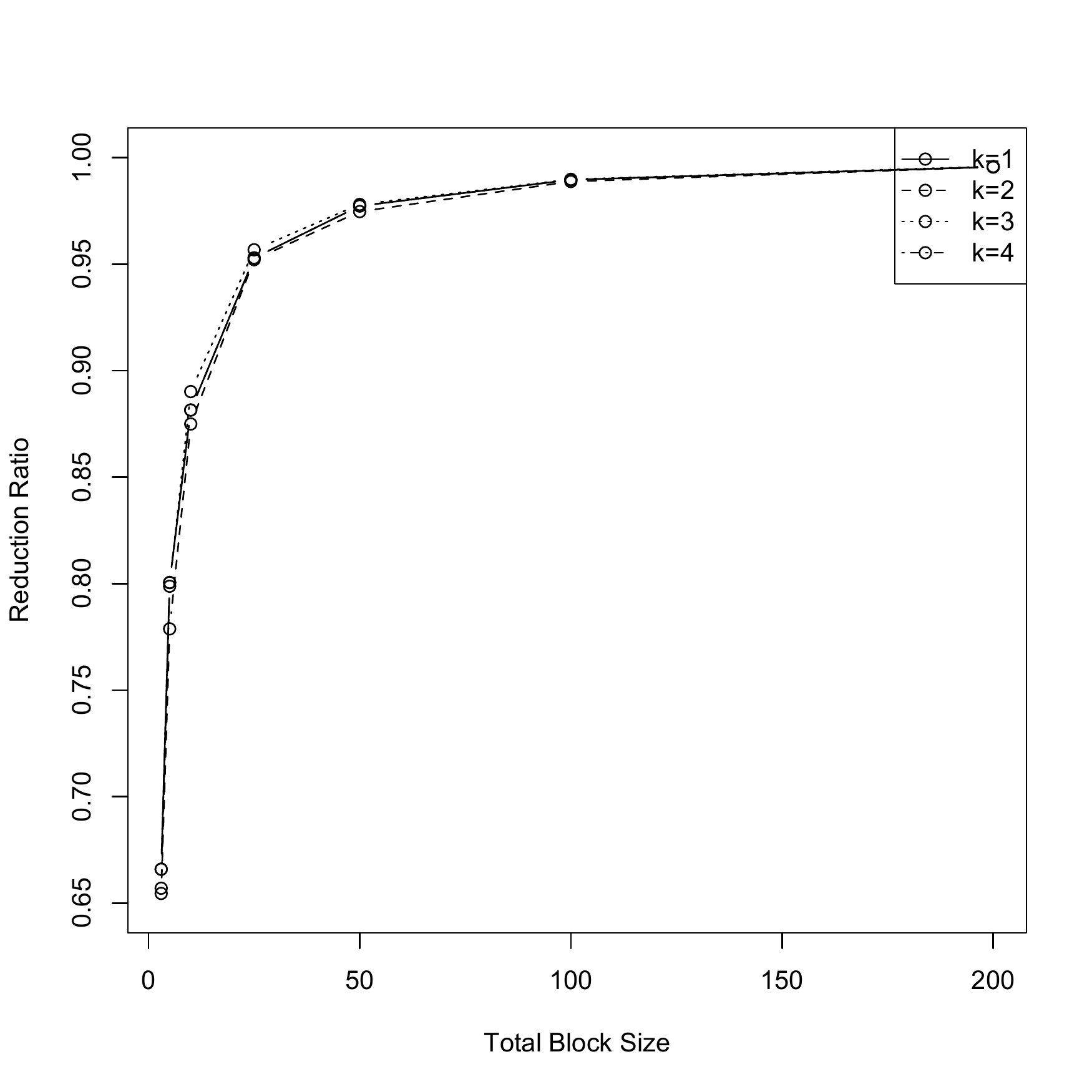}
\caption{\text{RLdata10000} dataset. Left:  For TLSH, we see the RR versus the number of shingles, where the RR is always very high. We emphasize that TLSH does about as well on the RR as any of the other methods, and certainly does much better than many traditional blocking methods and KNN. (The RR is always above $98\%$ for all shingles with $b=26$.)
Right: For KLSH, we illustrate the RR versus the total number of blocks for various $k=1,\ldots,4$ illustrating that as the number of blocks increases, the RR increases dramatically.  When the total block size is at least 25, the RR $\geq 95\%.$}
\label{reduction}
\end{figure}

\begin{figure}
\centering
\includegraphics[width=0.45\textwidth]{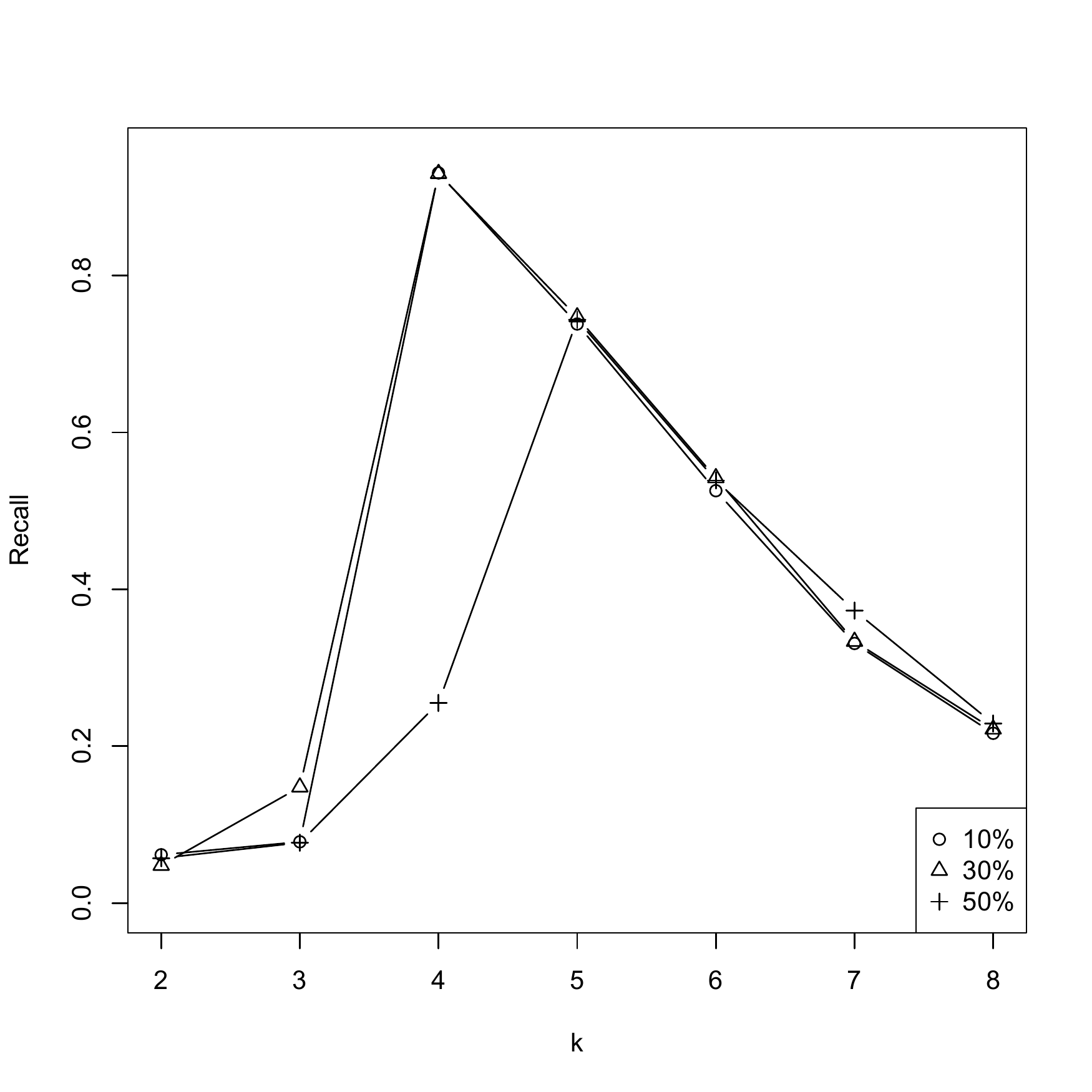}
\includegraphics[width=0.45\textwidth]{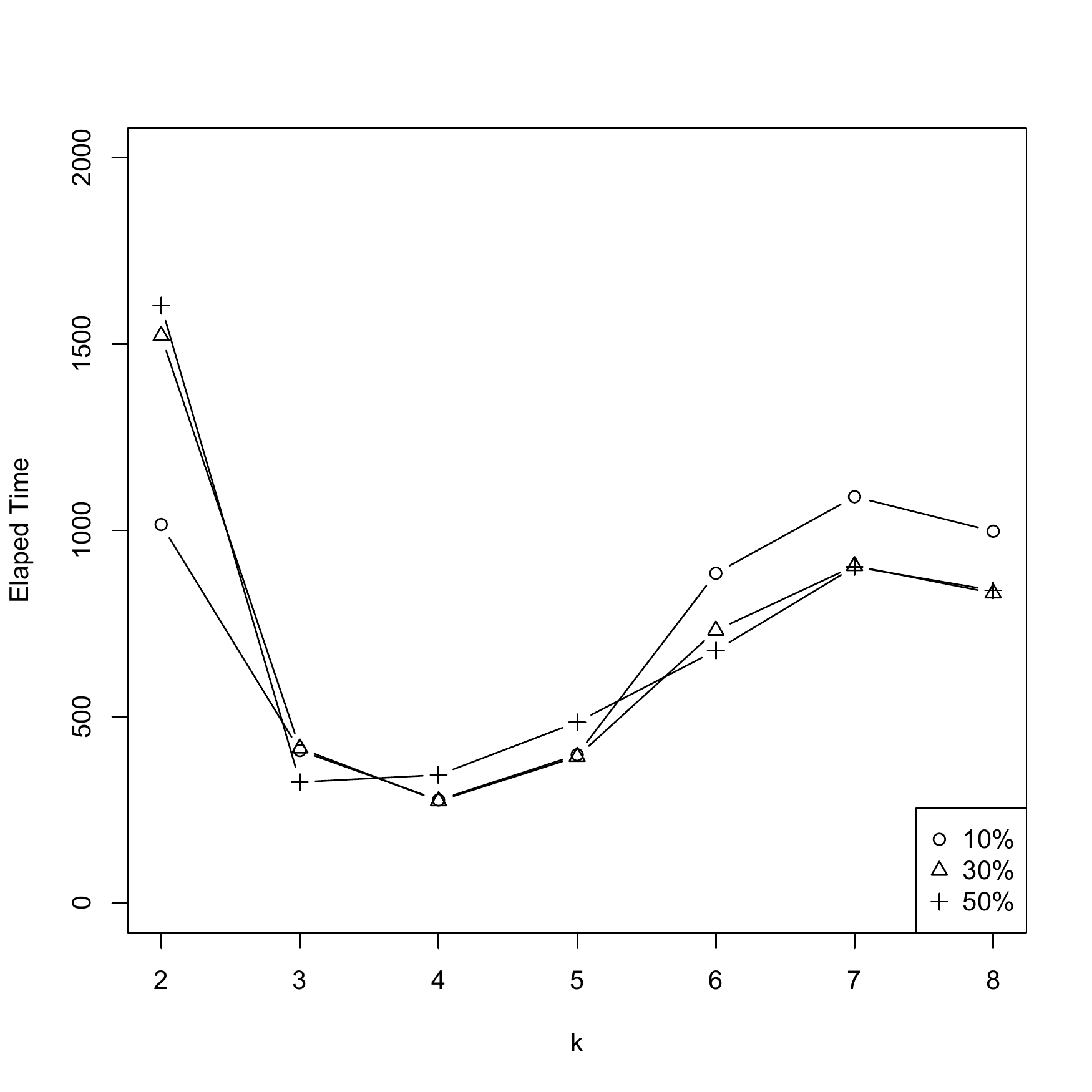}
\caption{ Left: We run TLSH for 10 percent duplicates, as before, the application is quite sensitive to $b,k.$ Hence, it is quite easy to find values of $b,k$ such that the recall is very low or if tuning is done properly, we can find values of $b,k$ where the recall is acceptable. We note this relies on very good ground truth. The only value of $k$ we recommend is 4 since it is close to 90\% recall. The computational time is the same as previously. Right: Elapsed time for 10, 30, and 50 percent duplicates on ``noisy" dataset. }
\label{reduction_new}
\end{figure}

\begin{figure}[t!]
\centering
\includegraphics[width=0.48\textwidth]{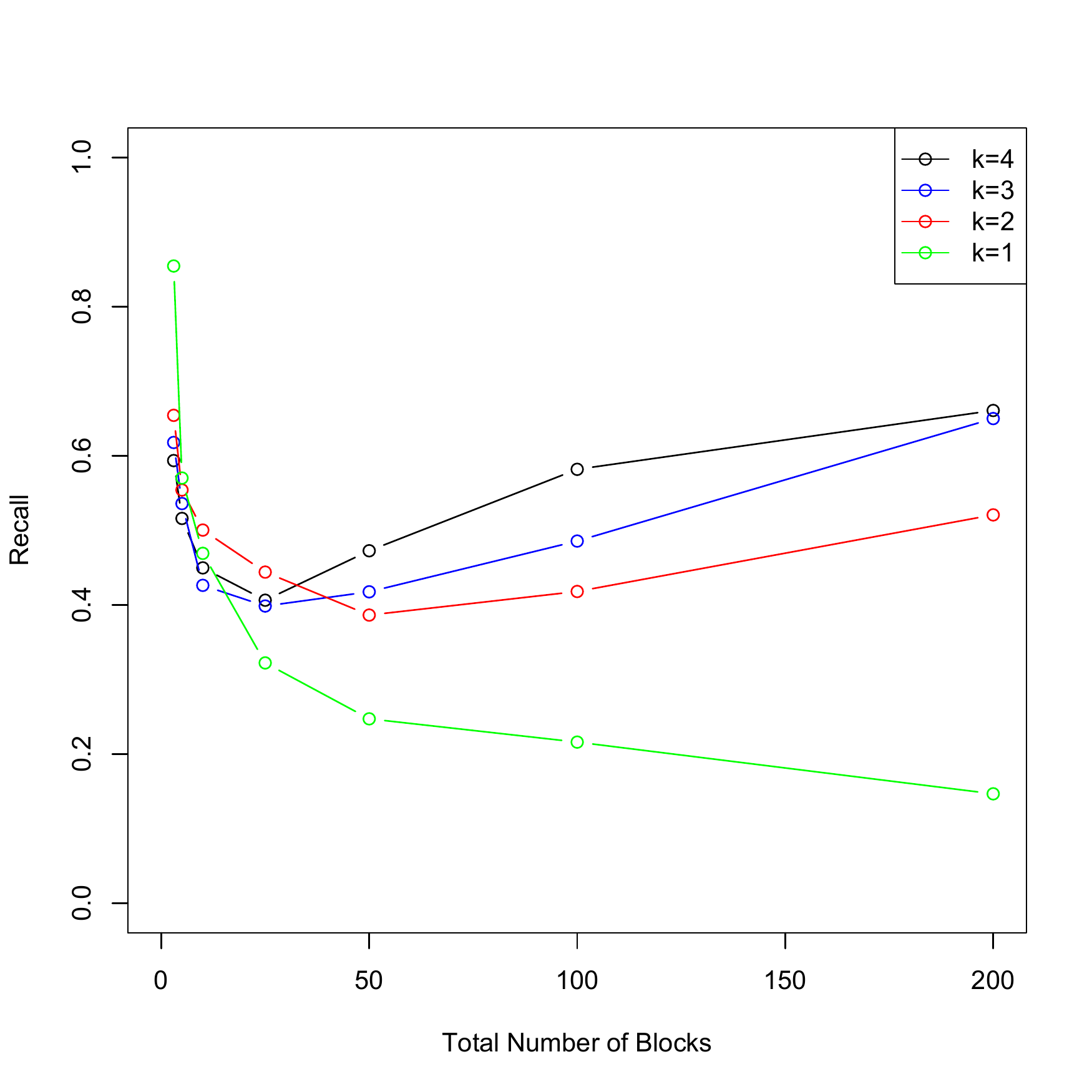}
\includegraphics[width=0.48\textwidth]{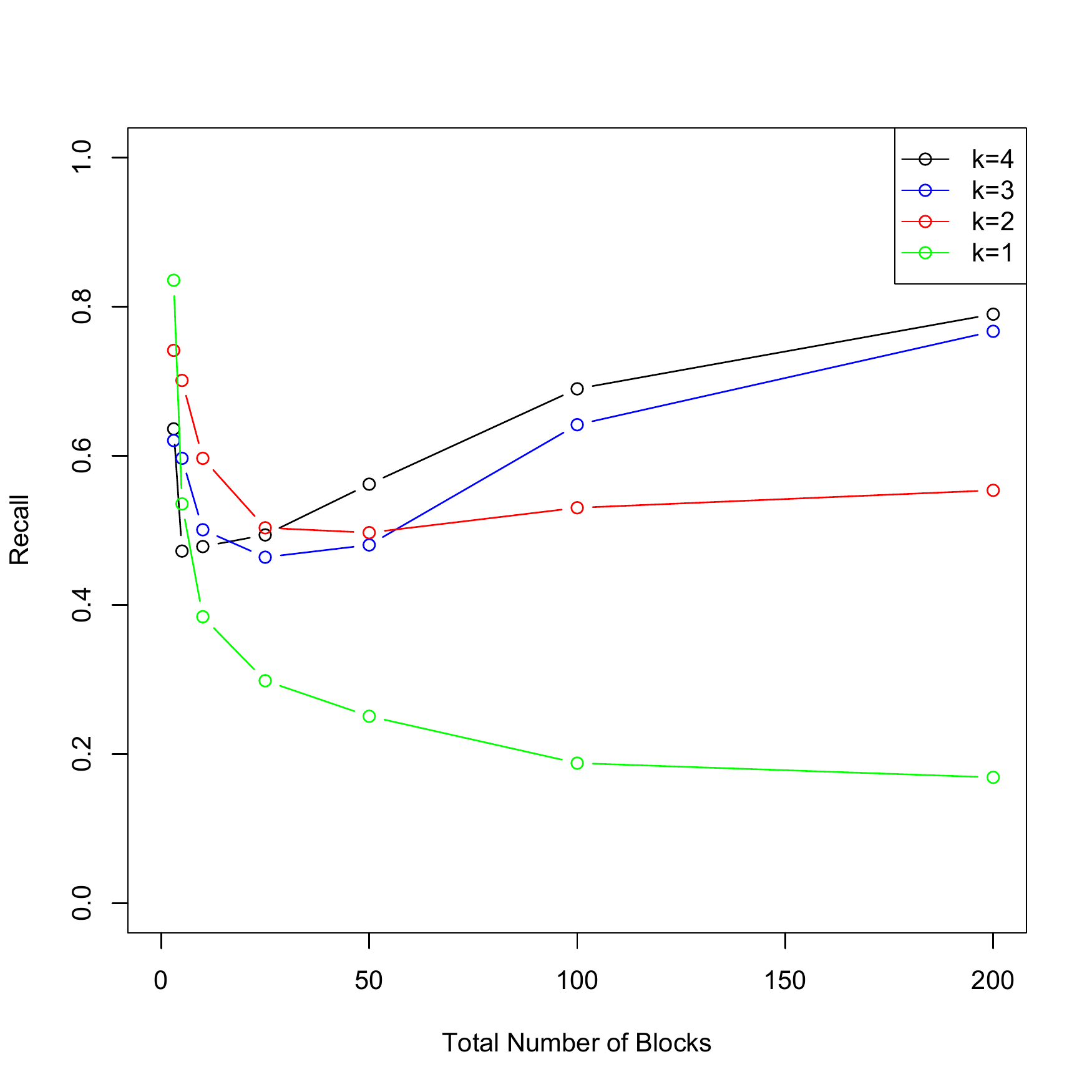}
\caption{ We run KLSH at 10 percent duplicates with p=100 (left) and p=150 (right). We see as the number of permutations increases (left figure), the recall increases. The behavior is the same for 30 and 50 percent duplicates. This indicates that KLSH needs to be tuned for each application based on $p.$ 
}
\label{klsh_recall_new}
\end{figure}

We analyze the \texttt{RLdata10000} database for TLSH and KLSH.  As we increase $k$ under TLSH, we see that the recall peaks at $k=5,$ and does very poorly (below 40\% recall) when $k\leq 4$.  For KLSH, the highest and most consistent recall is when $k=2,$ since it is always above 80\% and it is about the same no matter the total number of blocks chosen (see Figure \ref{distort_10000}). In terms of RR, we see that TLSH performs extremely poorly as the total number of blocks increases, whereas KLSH performs extremely well in terms of RR comparatively (Figure \ref{reduction}).  Figure \ref{comp_time} shows empirically that the running time for both KLSH and TLSH scales quadratically with the $n$, matching our asymptotic derivation. We then analyze the ``noisy" database for TLSH and KLSH (see Figures  \ref{reduction_new} and \ref{klsh_recall_new}).

\subsubsection{Comparisons of Methods}
In terms of comparing to the methods presented in Table \ref{t:naive_results}, we find that TLSH is not comparable in terms of recall or RR.  However, KLSH easily beats Criteria 1--2 and competes with Criteria 3--4 on both recall and RR. It does not perform as well in terms of recall as the rest of the criteria, however, it \emph{may} in other applications with more complex information for each record (this is a subject of future work). When comparing the Table \ref{t:naive_results2} to TLSH and KLSH when run for the noisy datafile, we find that TLSH and KLSH usually do better when tuned properly, however not always. Due to the way these files have been constructed, more investigation need to be done in terms of how naive methods work for real work type applications versus LSH-based methods. 

Comparing to other blocking methods, both KLSH and TLSH outperform KNN in terms of recall (and RR for the noisy datafiles).  We find that for this dataset, canopies do not perform well in terms of recall or RR unless a specific threshold $t_1$ is chosen.  However, given this choice of $t_1$, this approach yields either high recall and low RR or vice versa, making canopies undesirable according to our criteria.  

For the \texttt{RLdata10000} dataset, the simple yet effective traditional blocking methods and KLSH perform best in terms of balancing both high recall and high RR.  As already stated, we expect the performance of these to be heavily application-dependent.  Additionally, note that each method relies on high-quality labeled record linkage data to measure the recall and RR and the clustering methods require tuning parameters, which can be quite sensitive. Our studies show that TLSH is the least sensitive in general and further explorations should be done here.  Future work should explore the characteristics of the underlying datasets for which one method would be preferred over another.

\subsubsection{Sensitivity Analysis on \texttt{RLdata500} and \texttt{RLdata10000}}
A sensitivity analysis is given for KLSH and TLSH. 
For TLSH, the \texttt{RLdata500} dataset is not very sensitive to $b$ since the recall is always above 80\% whereas the \texttt{RLdata10000} dataset is quite sensitive to the band, and we recommend the use of a band of 21--22 since the recall for these $b$ is $\approx$ 96\%, although this may change for other datasets. We then evaluate TLSH using the ``best'' choice of the band for shingled values from $k=1,\ldots 5. $ The sensitivity analysis for the ``noisy" datafiles was quite similar to that described above, where a band of 22 was deemed the most appropriate for TLSH. For KLSH, we found that we needed to increase the number of permutations slightly to improve the recall and recommend $p=150.$

\begin{figure}
\centering
\includegraphics[width=0.45\textwidth]{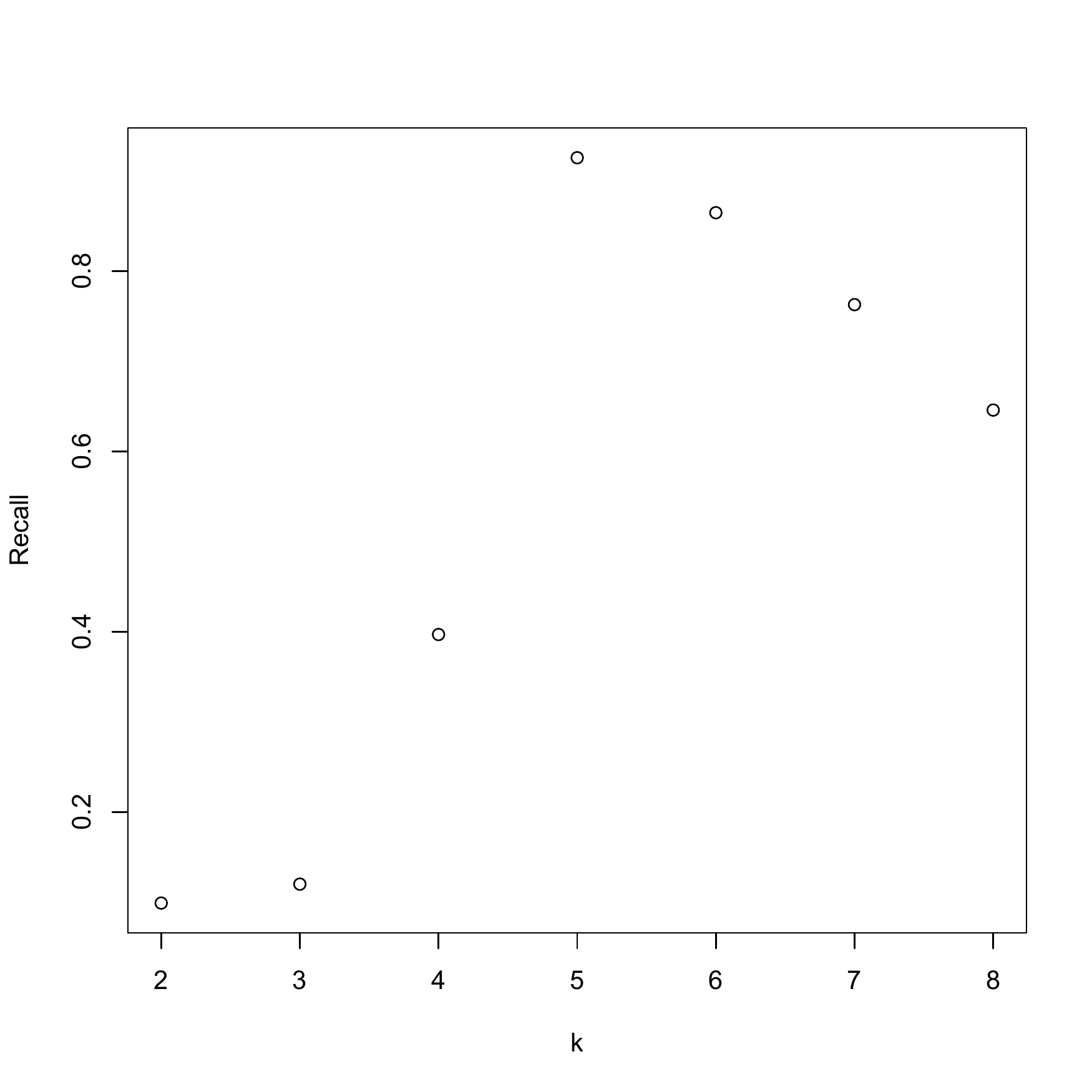}
\includegraphics[width=0.45\textwidth]{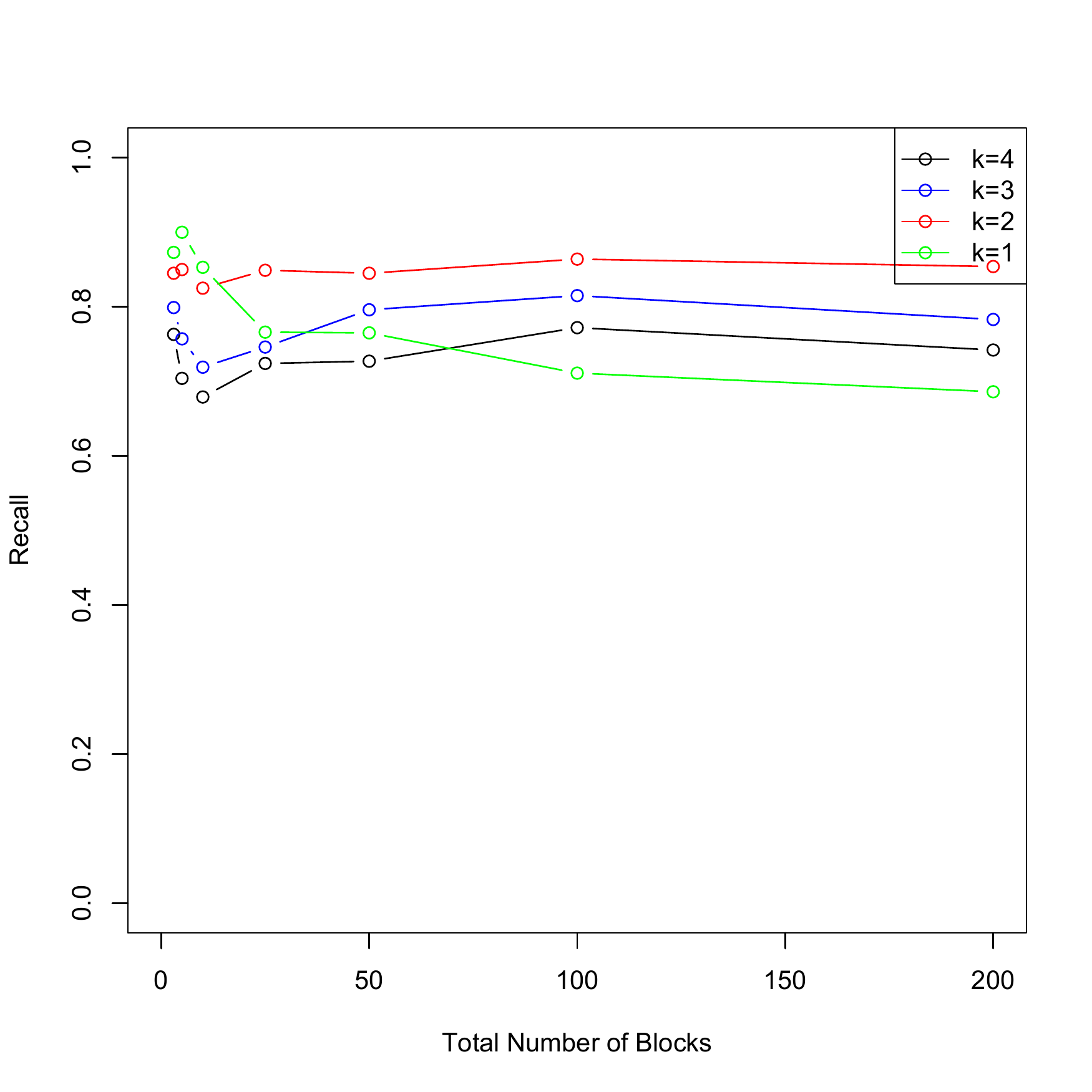}
\caption{\text{RLdata10000} dataset. Left: Recall versus number of shingles $k$ for KLSH. The highest recall occurs at $k=5$. Right: Recall versus the total number of blocks, where we vary the number of shingles $k$.  We find that the highest recall is for $k=2$.}
\label{distort_10000}
\end{figure}

For KLSH, we find that when the number of random permutations $p$ is above 100, the recall does not change considerably. We refer back to Figure \ref{distort_10000} (right), which illustrates the recall versus number of blocks when $p = 100.$  When $k=4,$ the recall is always above 70\%.  However, we find that when $k=2,$ the recall is always above 80\%.

\section{Discussion}
\label{sec:disc} 

We have explored two LSH methods for blocking, one of which would naturally fit into the privacy preserving record linkage (PPRL) framework, since the method could be made to be private by creating reference values for each individual in the database. This has been done for many blocking methods in the context of PPRL \citep{durham_2012, vatsalan_2011,karakasidis_2012, kuzu_2011}.  KLSH performs just as well or better than commonly used blocking methods, such as some simple traditional blocking methods, nearest neighbor clustering approaches, and canopies \citep{vatsalan_2013, mccallum_2000}. One drawback is that like LSH-based methods, it must be tuned for each application since it is sensitive to the tuning parameters. Thus, some \emph{reliable} training data must be available  to evaluate the recall and RR (and tune KLSH or clustering type methods).  In many situations, a researcher may be better off by using domain-specific knowledge to reduce the set of comparisons, as shown in \S \ref{ss:naive_results}.

LSH-methods have been described elsewhere as ``private blocking" due to the hashing step.  However, they do not in fact provide any formal privacy guarantees in our setting.  The new variant that we have introduced, KLSH, does satisfy the $k$-anonymity criterion for the de-duplication of a single file.  However, the data remain subject to intruder attacks, as the literature on differential privacy makes clear, and the vulnerability is greater the smaller the value of $k$.   Our broader goal, however, is to merge and analyze data from multiple files.  Privacy protection in that context is far more complicated.  Even if one could provide privacy guarantees for each file separately, it would still be possible to identify specific entities or  \emph{sensitive} information regarding entities in the merged database.  

The approach of PPRL reviewed in \cite{hall12} sets out to deal with this problem. Merging data from multiple files with the same or similar values without releasing their attributes is what PPRL hopes to achieve.  Indeed, one of course needs to go further, since performing statistical analyses on the merged database is the real objective of PPRL.   Whether the new ``private blocking" approaches discussed offer any progress on this problem, it is unclear at best.  Adequately addressing the PPRL goals remains elusive, as do formal privacy guarantees, be they from differential privacy or other methods. 



%

%

\section*{Acknowledgements}
We  thank Peter Christen, Patrick Ball, and Cosma Shalizi for thoughtful conversations that led to to early versions of this manuscript. We also thank the reviewers for their suggestions and comments.

\clearpage
\newpage

\bibliographystyle{ims}
\bibliography{chomp}

\begin{thebibliography}{23}
\expandafter\ifx\csname natexlab\endcsname\relax\def\natexlab#1{#1}\fi
\expandafter\ifx\csname url\endcsname\relax
  \def\url#1{\texttt{#1}}\fi
\expandafter\ifx\csname urlprefix\endcsname\relax\def\urlprefix{URL }\fi
\providecommand{\eprint}[2][]{\url{#2}}

\bibitem[{Christen(2005)}]{Christen05}
\textsc{Christen, P.} (2005).
\newblock {Probabilistic Data Generation for Deduplication and Data Linkage}.
\newblock In \textit{Proceedings of the Sixth International Conference on
  Intelligent Data Engineering and Automated Learning (IDEAL'05)}. 109--116.

\bibitem[{Christen(2012)}]{christen_2011}
\textsc{Christen, P.} (2012).
\newblock A survey of indexing techniques for scalable record linkage and
  deduplication.
\newblock \textit{IEEE Transactions on Knowledge and Data Engineering},
  \textbf{24}.

\bibitem[{Christen and Pudjijono(2009{\natexlab{a}})}]{christen_2009}
\textsc{Christen, P.} and \textsc{Pudjijono, A.} (2009{\natexlab{a}}).
\newblock Accurate synthetic generation of realistic personal information.
\newblock In \textit{Advances in Knowledge Discovery and Data Mining}.
  Springer, 507--514.

\bibitem[{Christen and Pudjijono(2009{\natexlab{b}})}]{ChristenPudjijono09}
\textsc{Christen, P.} and \textsc{Pudjijono, A.} (2009{\natexlab{b}}).
\newblock {Accurate Synthetic Generation of Realistic Personal Information}.
\newblock In \textit{Advances in Knowledge Discovery and Data Mining}
  (T.~Theeramunkong, B.~Kijsirikul, N.~Cercone and T.-B. Ho, eds.), vol. 5476
  of \textit{Lecture Notes in Computer Science}. Springer Berlin Heidelberg,
  507--514.
\newblock \urlprefix\url{http://dx.doi.org/10.1007/978-3-642-01307-2_47}.

\bibitem[{Christen and Vatsalan(2013)}]{ChristenVatsalan13}
\textsc{Christen, P.} and \textsc{Vatsalan, D.} (2013).
\newblock {Flexible and Extensible Generation and Corruption of Personal Data}.
\newblock In \textit{Proceedings of the ACM International Conference on
  Information and Knowledge Management (CIKM 2013)}.

\bibitem[{Clauset et~al.(2004)Clauset, Newman and Moore}]{clauset_2004}
\textsc{Clauset, A.}, \textsc{Newman, M.~E.} and \textsc{Moore, C.} (2004).
\newblock Finding community structure in very large networks.
\newblock \textit{Physical review E}, \textbf{70} 066111.

\bibitem[{Durham(2012)}]{durham_2012}
\textsc{Durham, E.~A.} (2012).
\newblock \textit{A framework for accurate, efficient private record linkage}.
\newblock Ph.D. thesis, Vanderbilt University.

\bibitem[{Fortunato(2010)}]{fortunato_2010}
\textsc{Fortunato, S.} (2010).
\newblock Community detection in graphs.
\newblock \textit{Physics Reports}, \textbf{486} 75--174.

\bibitem[{Goldenberg et~al.(2010)Goldenberg, Zheng, Fienberg and
  Airoldi}]{goldenberg_2010}
\textsc{Goldenberg, A.}, \textsc{Zheng, A.~X.}, \textsc{Fienberg, S.~E.} and
  \textsc{Airoldi, E.~M.} (2010).
\newblock A survey of statistical network models.
\newblock \textit{Foundations and Trends{\textregistered} in Machine Learning},
  \textbf{2} 129--233.

\bibitem[{Hall and Fienberg(2012)}]{hall12}
\textsc{Hall, R.} and \textsc{Fienberg, S.} (2012).
\newblock Valid statistical inference on automatically matched files.
\newblock In \textit{Privacy in Statistical Databases 2012} (J.~Domingo-Ferrer
  and I.~Tinnirello, eds.), vol. 7556 of \textit{Lecture Notes in Computer
  Science}. Springer, Berlin, 131--142.

\bibitem[{Herzog et~al.(2007)Herzog, Scheuren and Winkler}]{Herzog_2007}
\textsc{Herzog, T.}, \textsc{Scheuren, F.} and \textsc{Winkler, W.} (2007).
\newblock \textit{Data Quality and Record Linkage Techniques}.
\newblock Springer, New York.

\bibitem[{Herzog et~al.(2010)Herzog, Scheuren and Winkler}]{Herzog:2010}
\textsc{Herzog, T.}, \textsc{Scheuren, F.} and \textsc{Winkler, W.} (2010).
\newblock Record linkage.
\newblock \textit{Wiley Interdisciplinary Reviews: Computational Statistics},
  \textbf{2} DOI: 10.1002/wics.108.

\bibitem[{Karakasidis and Verykios(2012)}]{karakasidis_2012}
\textsc{Karakasidis, A.} and \textsc{Verykios, V.~S.} (2012).
\newblock Reference table based k-anonymous private blocking.
\newblock In \textit{Proceedings of the 27th Annual ACM Symposium on Applied
  Computing}. ACM, 859--864.

\bibitem[{Kuzu et~al.(2011)Kuzu, Kantarcioglu, Durham and Malin}]{kuzu_2011}
\textsc{Kuzu, M.}, \textsc{Kantarcioglu, M.}, \textsc{Durham, E.} and
  \textsc{Malin, B.} (2011).
\newblock A constraint satisfaction cryptanalysis of bloom filters in private
  record linkage.
\newblock In \textit{Privacy Enhancing Technologies}. Springer, 226--245.

\bibitem[{Liang et~al.(2014)Liang, Wang, Christen and Gayler}]{christen_2014}
\textsc{Liang, H.}, \textsc{Wang, Y.}, \textsc{Christen, P.} and
  \textsc{Gayler, R.} (2014).
\newblock Noise-tolerant approximate blocking for dynamic real-time entity
  resolution.
\newblock In \textit{Eighteenth Pacific-Asia Conference on Knowledge Discovery
  and Data Mining (PAKDD14)} (Z.-H. Zhou, A.~L.~P. Chen, V.~S. Tseng and B.~H.
  Tu, eds.). Springer, New York, NY, forthcoming.

\bibitem[{McCallum et~al.(2000)McCallum, Nigam and Ungar}]{mccallum_2000}
\textsc{McCallum, A.}, \textsc{Nigam, K.} and \textsc{Ungar, L.~H.} (2000).
\newblock Efficient clustering of high-dimensional data sets with application
  to reference matching.
\newblock In \textit{Proceedings of the sixth ACM SIGKDD international
  conference on Knowledge discovery and data mining}. ACM, 169--178.

\bibitem[{Pasula et~al.(2003)Pasula, Marthi, Milch, Russell and
  Shpitser}]{pasula_2003}
\textsc{Pasula, H.}, \textsc{Marthi, B.}, \textsc{Milch, B.}, \textsc{Russell,
  S.} and \textsc{Shpitser, I.} (2003).
\newblock Identity uncertainty and citation matching.
\newblock \textit{Advances in neural information processing systems}
  1425--1432.

\bibitem[{Paulev{\'e} et~al.(2010)Paulev{\'e}, J{\'e}gou and
  Amsaleg}]{pauleve_2010}
\textsc{Paulev{\'e}, L.}, \textsc{J{\'e}gou, H.} and \textsc{Amsaleg, L.}
  (2010).
\newblock Locality sensitive hashing: A comparison of hash function types and
  querying mechanisms.
\newblock \textit{Pattern Recognition Letters}, \textbf{31} 1348--1358.

\bibitem[{Rajaraman and Ullman(2012)}]{rajaraman_2012}
\textsc{Rajaraman, A.} and \textsc{Ullman, J.~D.} (2012).
\newblock \textit{Mining of massive datasets}.
\newblock Cambridge University Press.

\bibitem[{Steorts et~al.(2013)Steorts, Hall and Fienberg}]{steorts_2013b}
\textsc{Steorts, R.~C.}, \textsc{Hall, R.} and \textsc{Fienberg, S.} (2013).
\newblock A {B}ayesian approach to graphical record linkage and de-duplication.
\newblock \textit{Submitted}.

\bibitem[{Vatsalan and Christen(2013)}]{vatsalan_2013}
\textsc{Vatsalan, D.} and \textsc{Christen, P.} (2013).
\newblock Sorted nearest neighborhood clustering for efficient private
  blocking.
\newblock In \textit{Advances in Knowledge Discovery and Data Mining}.
  Springer, 341--352.

\bibitem[{Vatsalan et~al.(2011)Vatsalan, Christen and Verykios}]{vatsalan_2011}
\textsc{Vatsalan, D.}, \textsc{Christen, P.} and \textsc{Verykios, V.~S.}
  (2011).
\newblock An efficient two-party protocol for approximate matching in private
  record linkage.
\newblock In \textit{Proceedings of the Ninth Australasian Data Mining
  Conference-Volume 121}. Australian Computer Society, Inc., 125--136.

\bibitem[{Winkler et~al.(2010)Winkler, Yancey and Porter}]{WYP:2010}
\textsc{Winkler, W.}, \textsc{Yancey, W.} and \textsc{Porter, E.} (2010).
\newblock Fast record linkage of very large files in support of decennial and
  administrative records projects.
\newblock In \textit{Proceedings of American Statistical Association Section on
  Survey Research Methods}.

\end{thebibliography}

\clearpage
\newpage
\normalsize
\appendix
\section*{Algorithms for KLSH and TLSH}
\label{sec:app}
We provide the algorithms for KLSH and TLSH below (see \S \ref{subsec:tlsh} and \ref{subsec:klsh}):

\begin{algorithm}[h]
  \DontPrintSemicolon
  Place similar records into blocks and impose transitivity\;
  \BlankLine
  \KwData{$X_{ij}$, tuning parameters $b, t,k$}
  \BlankLine
  Shingle each $X_{ij}$ into length-$k$ strings\;
  Create a binary matrix $M$ indicating which tokens appear in which records\;
  Create an integer-valued matrix $M^{\prime}$ of minhash signatures from $M$\;
  Divide the rows of $M^{\prime}$ into $b$ bands\;
  \For{each band} {
  $\;$ Apply a random hash function to the band of $M^{\prime}$\;
  $\;$ Record an edge between two records if the hash maps them to the same bucket\;
  }
  \While{the largest community has $> t$ records} {
    Cut the edge graph into finer communities using the algorithm of \cite{clauset_2004}
  }
  \Return{the final list of communities}
  \caption{Transitive Locality Sensitive Hashing (TLSH)}
  \label{alg:tlsh}
\end{algorithm}

\begin{algorithm}[T!]
  \DontPrintSemicolon
  Place similar records into blocks and using k-means clustering and random projections\;
  \BlankLine
  \KwData{$X_{ij}$, number of desired blocks $c$, tokenization tuning parameters $\tau$, number of projections $p$}
  \BlankLine
  \For{each record $X_{ij}$} {
    $\;$ \KwSet $v_{ij} = \textsc{Tokenize}(X_{ij}, \tau)$
  }
  \For{each token $w$}  {
    $\;$ \KwSet $N_w = $number of bags containing $w$ \;
    $\;$ \KwSet $IDF_w = \log{n/N_w}$ \;
    }
 \For{$m$ from $1$ to $p$} {
   $\;$ \KwSet $u_m = $ a random unit vector \;
   $\;$ \For{each bag-of-tokens vector $v_{ij}$}{
   $\;$ \KwSet $r_{ijm} = \sum_{w}{u_{iw} v_{ijw} IDF_w}$ \;
   } \;
 }
 \Return{\textsc{KMEANS}($r$,$c$)}\;
  \caption{K-Means Locality Sensitive Hashing (KLSH).  {\scriptsize The number of blocks $c$ is set by $c= n/(\text{desired avg. number of records per block}).$} }
  \label{alg:tlsh}
\end{algorithm}

\end{document}